\documentclass[conference]{IEEEtran}
\IEEEoverridecommandlockouts
\usepackage{amsmath,amsfonts}
\usepackage{algpseudocode}
\usepackage{algorithm}
\usepackage{amsmath,amsfonts}
\usepackage{algpseudocode}
\usepackage{algorithm}
\usepackage{array}
\usepackage[caption=false,font=normalsize,labelfont=sf,textfont=sf]{subfig}
\usepackage{textcomp}
\usepackage{stfloats}
\usepackage{url}
\usepackage{verbatim}
\usepackage{color,soul}
\usepackage{graphicx}
\usepackage{hyperref}
\usepackage{cite}
\usepackage[nolist]{acronym}
\usepackage{xcolor,cite,etoolbox}
\usepackage{amssymb}
\usepackage{enumitem}

\acrodefplural{HPA}[HPAs]{high power amplifiers}

\begin{acronym}
	\acro{AP}{access point}
	\acro{AN}{artificial noise}
	\acro{SI}{self interference}
    \acro{MIMO}{multiple-input, multiple-output}
    \acro{AWGN}{additive white Gaussian noise}
    \acro{KPI}{key performance indicator}
      \acro{KPIs}{key performance indicators}  
    \acro{D2D}{device-to-device}
    \acro{ISAC}{integrated sensing and communications}
    \acro{DFRC}{dual-functional radar and communication}
    \acro{AoA}{angle of arrival}
    \acro{AoAs}{angles of arrival}
    \acro{MU-MIMO}{multi-user, multiple-input, multiple-output}
    \acro{ToA}{time of arrival}
    \acro{PCA}{principle component analysis}
    \acro{EVM}{error vector magnitude}
    \acro{CPI}{coherent processing interval}
    \acro{eMBB}{enhanced mobile broadband}
    \acro{URLLC}{ultra-reliable low latency communications}
    \acro{mMTC}{massive machine type communications}
	\acro{QCQP}{quadratic constrained quadratic programming}
	\acro{BnB}{branch and bound}
	\acro{SER}{symbol error rate}
	\acro{LFM}{linear frequency modulation}
	\acro{BS}{base station}
	\acro{SLNR}{signal-to-leakage-plus-noise ratio}
	\acro{UE}{user equipment}
	\acro{DL}{deep learning}
	\acro{CS}{compressed sensing}
	\acro{PR}{passive radar}
	\acro{LMS}{least mean squares}
	\acro{NLMS}{normalized least mean squares}
	\acro{ZF}{zero forcing}
	\acro{RIS}{reconfigurable intelligent surface}
	\acro{RISs}{reconfigurable intelligent surfaces}
	\acro{IRS}{intelligent reflecting surface}
	\acro{IRSs}{intelligent reflecting surfaces}
	\acro{NOMA}{non-orthogonal multiple access}
	\acro{TWRN}{two-way relay networks}
	\acro{LISA}{large intelligent surface/antennas}
	\acro{LISs}{large intelligent surfaces}
	\acro{OFDM}{orthogonal frequency-division multiplexing}
	\acro{EM}{electromagnetic}
	\acro{ISMR}{integrated sidelobe to mainlobe ratio}
	\acro{MSE}{mean squared error}
	\acro{SNR}{signal-to-noise ratio}
	\acro{IoT}{internet of things}
	\acro{SRP}{successful recovery probability}
	\acro{CDF}{cumulative distribution function}
	\acro{ULA}{uniform linear array}
	\acro{RSS}{received signal strength}
	\acro{SU}{single user}
	\acro{MU}{multi-user}
	\acro{CSI}{channel state information}
	\acro{OFDM}{orthogonal frequency-division multiplexing}
	\acro{DL}{downlink}
	\acro{i.i.d}{independently and identically distributed}
	\acro{UL}{uplink}
	\acro{LoS}{line of sight}
	\acro{DFT}{discrete Fourier transform}
    \acro{HPA}{high power amplifier}
    \acro{IBO}{input power back-off}
    \acro{MIMO}{multiple-input, multiple-output}
    \acro{PAPR}{peak-to-average power ratio}
    \acro{AWGN}{additive white Gaussian noise}
    \acro{KPI}{key performance indicator}
    \acro{FD}{full duplex}
    \acro{KPIs}{key performance indicators}  
    \acro{D2D}{device-to-device}
    \acro{ISAC}{integrated sensing and communications}
    \acro{DFRC}{dual-functional radar and communication}
    \acro{AoA}{angle of arrival}
    \acro{AoD}{angle of departure}
    \acro{ToA}{time of arrival}
    \acro{EVM}{error vector magnitude}
    \acro{CPI}{coherent processing interval}
    \acro{eMBB}{enhanced mobile broadband}
    \acro{URLLC}{ultra-reliable low latency communications}
    \acro{mMTC}{massive machine type communications}
	\acro{QCQP}{quadratic constrained quadratic programming}
	\acro{BnB}{branch and bound}
	\acro{SER}{symbol error rate}
	\acro{LFM}{linear frequency modulation}
	\acro{ADMM}{alternating direction method of multipliers}
	\acro{CCDF}{complementary cumulative distribution function}
	\acro{MISO}{multiple-input, single-output}
	\acro{CSI}{channel state information}
	\acro{LDPC}{low-density parity-check}
	\acro{BCC}{binary convolutional coding}
	\acro{IEEE}{Institute of Electrical and Electronics Engineers}
	\acro{ULA}{uniform linear antenna}
	\acro{B5G}{beyond 5G}
	\acro{MOOP}{multi-objective optimization problem}
	\acro{ML}{maximum likelihood}
	\acro{FML}{fused maximum likelihood}
	\acro{RF}{radio frequency}
	\acro{AM/AM}{amplitude modulation/amplitude modulation}
	\acro{AM/PM}{amplitude modulation/phase modulation}
	\acro{BS}{base station}
	\acro{SCA}{successive convex approximation}
	\acro{MM}{majorization-minimization}
	\acro{LNCA}{$\ell$-norm cyclic algorithm}
	\acro{OCDM}{orthogonal chirp-division multiplexing}
	\acro{TR}{tone reservation}
	\acro{COCS}{consecutive ordered cyclic shifts}
	\acro{LS}{least squares}
	\acro{CVE}{coefficient of variation of envelopes}
	\acro{BSUM}{block successive upper-bound minimization}
	\acro{ICE}{iterative convex enhancement}
	\acro{SOA}{sequential optimization algorithm}
	\acro{BCD}{block coordinate descent}
	\acro{SINR}{signal to interference plus noise ratio}
	\acro{MICF}{modified iterative clipping and filtering}
	\acro{PSL}{peak side-lobe level}
	\acro{SDR}{semi-definite relaxation}
	\acro{KL}{Kullback-Leibler}
	\acro{ADSRP}{alternating direction sequential relaxation programming}
	\acro{MUSIC}{MUltiple SIgnal Classification}
	\acro{EVD}{eigenvalue decomposition}
	\acro{SVD}{singular value decomposition}
	\acro{LDGM}{low-density generator matrix}
	\acro{SAC}{sensing and communication}
	\acro{AN}{artificial noise}
	\acro{SI}{self interference}
    \acro{MIMO}{multiple-input, multiple-output}
    \acro{AWGN}{additive white Gaussian noise}
    \acro{KPI}{key performance indicator}
      \acro{KPIs}{key performance indicators}  
    \acro{D2D}{device-to-device}
    \acro{RCS}{radar cross-section}
    \acro{ISAC}{integrated sensing and communication}
    \acro{DFRC}{dual-functional radar and communication}
    \acro{AoA}{angle of arrival}
    \acro{AoAs}{angles of arrival}
    \acro{ToA}{time of arrival}
    \acro{EVM}{error vector magnitude}
    \acro{CPI}{coherent processing interval}
    \acro{eMBB}{enhanced mobile broadband}
    \acro{URLLC}{ultra-reliable low latency communications}
    \acro{mMTC}{massive machine type communications}
	\acro{QCQP}{quadratic constrained quadratic programming}
	\acro{BnB}{branch and bound}
	\acro{SER}{symbol error rate}
	\acro{LFM}{linear frequency modulation}
	\acro{BS}{base station}
	\acro{UE}{user equipment}
	\acro{DL}{deep learning}
	\acro{CS}{compressed sensing}
	\acro{PR}{passive radar}
	\acro{LMS}{least mean squares}
	\acro{NLMS}{normalized least mean squares}
	\acro{RIS}{reconfigurable intelligent surface}
	\acro{RISs}{reconfigurable intelligent surfaces}
	\acro{IRS}{intelligent reflecting surface}
	\acro{IRSs}{intelligent reflecting surfaces}
	\acro{LISA}{large intelligent surface/antennas}
	\acro{LISs}{large intelligent surfaces}
	\acro{OFDM}{orthogonal frequency-division multiplexing}
	\acro{EM}{electromagnetic}
	\acro{ISMR}{integrated sidelobe to mainlobe ratio}
	\acro{MSE}{mean squared error}
	\acro{SNR}{signal-to-noise ratio}
	\acro{SRP}{successful recovery probability}
	\acro{CDF}{cumulative distribution function}
	\acro{ULA}{uniform linear array}
	\acro{RSS}{received signal strength}
	\acro{SU}{single user}
	\acro{MU}{multi-user}
	\acro{CSI}{channel state information}
	\acro{OFDM}{orthogonal frequency-division multiplexing}
	\acro{DL}{downlink}
	\acro{i.i.d}{independently and identically distributed}
	\acro{UL}{uplink}
	\acro{LoS}{line of sight}
	\acro{DFT}{discrete Fourier transform}
    \acro{HPA}{high power amplifier}
    \acro{IBO}{input power back-off}
    \acro{MIMO}{multiple-input, multiple-output}
    \acro{PAPR}{peak-to-average power ratio}
    \acro{AWGN}{additive white Gaussian noise}
    \acro{KPI}{key performance indicator}
    \acro{FD}{full duplex}
    \acro{KPIs}{key performance indicators}  
    \acro{D2D}{device-to-device}
    \acro{ISAC}{integrated sensing and communication}
    \acro{DFRC}{dual-functional radar and communication}
    \acro{AoA}{angle of arrival}
    \acro{AoD}{angle of departure}
    \acro{ToA}{time of arrival}
    \acro{EVM}{error vector magnitude}
    \acro{CPI}{coherent processing interval}
    \acro{eMBB}{enhanced mobile broadband}
    \acro{URLLC}{ultra-reliable low latency communications}
    \acro{mMTC}{massive machine type communications}
	\acro{QCQP}{quadratic constrained quadratic programming}
	\acro{BnB}{branch and bound}
	\acro{SER}{symbol error rate}
	\acro{LFM}{linear frequency modulation}
	\acro{ADMM}{alternating direction method of multipliers}
	\acro{CCDF}{complementary cumulative distribution function}
	\acro{MISO}{multiple-input, single-output}
	\acro{CSI}{channel state information}
	\acro{LDPC}{low-density parity-check}
	\acro{BCC}{binary convolutional coding}
	\acro{IEEE}{Institute of Electrical and Electronics Engineers}
	\acro{ULA}{uniform linear antenna}
	\acro{B5G}{beyond 5G}
	\acro{MOOP}{multi-objective optimization problem}
	\acro{ML}{maximum likelihood}
	\acro{FML}{fused maximum likelihood}
	\acro{RF}{radio frequency}
	\acro{AM/AM}{amplitude modulation/amplitude modulation}
	\acro{AM/PM}{amplitude modulation/phase modulation}
	\acro{BS}{base station}
	\acro{SCA}{successive convex approximation}
	\acro{MM}{majorization-minimization}
	\acro{LNCA}{$\ell$-norm cyclic algorithm}
	\acro{OCDM}{orthogonal chirp-division multiplexing}
	\acro{TR}{tone reservation}
	\acro{COCS}{consecutive ordered cyclic shifts}
	\acro{LS}{least squares}
	\acro{CVE}{coefficient of variation of envelopes}
	\acro{BSUM}{block successive upper-bound minimization}
	\acro{ICE}{iterative convex enhancement}
	\acro{SOA}{sequential optimization algorithm}
	\acro{BCD}{block coordinate descent}
	\acro{SINR}{signal to interference plus noise ratio}
	\acro{MICF}{modified iterative clipping and filtering}
	\acro{PSL}{peak side-lobe level}
	\acro{SDR}{semi-definite relaxation}
	\acro{XR}{extended reality}
	\acro{DoF}{degrees of freedom}
	\acro{KL}{Kullback-Leibler}
	\acro{mmWave}{milli-meter wave}
	\acro{ADSRP}{alternating direction sequential relaxation programming}
	\acro{MUSIC}{MUltiple SIgnal Classification}
	\acro{EVD}{eigenvalue decomposition}
	\acro{SVD}{singular value decomposition}
	\acro{i.i.d}{independent and identically distributed}
	\acro{PDF}{probability density function}
	\acro{GMM}{Gaussian mixture model}
	\acro{MI}{mutual information}
	\acro{MOOP}{multi-objective optimization problem}
	\acro{RSC}{restricted strongly convex}
	\acro{RSS}{restricted strongly smooth}
	\acro{ROC}{receiver operating characteristic}
	\acro{NMSE}{normalized mean square error}
	\acro{MMSE}{minimum mean square error}
	\acro{SER}{symbol error rate}
	\acro{QAM}{quadrature amplitude modulation}
	\acro{ZF}{zero-forcing}
	\acro{B5G}{beyond 5G}
	\acro{IoT}{internet of things}
	\acro{UAV}{unmanned aerial vehicle}
	\acro{DAM}{delay alignment modulation}
	\acro{ISR}{integrated sidelobe ratio}
	\acro{CRB}{Cram\'er-Rao bound}
	\acro{RCC}{radar-communication coexistence}
	\acro{NSP}{null space projection}
	\acro{OTFS}{orthogonal time frequency space}
	\acro{STARS}{simultaneously transmitting and reflecting surface}
	\acro{CRLB}{Cram\'er-Rao lower bound}
	\acro{PGD}{projected gradient descent}
    \acro{MIMO}{multiple-input, multiple-output}
    \acro{AWGN}{additive white Gaussian noise}
    \acro{KPI}{key performance indicator}
      \acro{KPIs}{key performance indicators}  
    \acro{D2D}{device-to-device}
    \acro{ISAC}{integrated sensing and communications}
    \acro{DFRC}{dual-functional radar and communication}
    \acro{AoA}{angle of arrival}
    \acro{AoAs}{angles of arrival}
    \acro{ToA}{time of arrival}
    \acro{EVM}{error vector magnitude}
    \acro{CPI}{coherent processing interval}
    \acro{eMBB}{enhanced mobile broadband}
    \acro{URLLC}{ultra-reliable low latency communications}
    \acro{mMTC}{massive machine type communications}
	\acro{QCQP}{quadratic constrained quadratic programming}
	\acro{BnB}{branch and bound}
	\acro{SER}{symbol error rate}
	\acro{LFM}{linear frequency modulation}
	\acro{BS}{base station}
	\acro{UE}{user equipment}
	\acro{DL}{deep learning}
	\acro{CS}{compressed sensing}
	\acro{PR}{passive radar}
	\acro{LMS}{least mean squares}
	\acro{NLMS}{normalized least mean squares}
	\acro{RIS}{reconfigurable intelligent surface}
	\acro{RISs}{reconfigurable intelligent surfaces}
	\acro{IRS}{intelligent reflecting surface}
	\acro{IRSs}{intelligent reflecting surfaces}
	\acro{LISA}{large intelligent surface/antennas}
	\acro{LISs}{large intelligent surfaces}
	\acro{OFDM}{orthogonal frequency-division multiplexing}
	\acro{EM}{electromagnetic}
	\acro{MSE}{mean squared error}
	\acro{SNR}{signal-to-noise ratio}
	\acro{SRP}{successful recovery probability}
	\acro{CDF}{cumulative distribution function}
	\acro{ULA}{uniform linear array}
	\acro{RSS}{received signal strength}
	\acro{ADC}{analog-to-digital converter}
	\acro{DR}{dynamic range}
\end{acronym}

\DeclareMathOperator{\SNR}{SNR}

\DeclareMathOperator{\dB}{dB}

\DeclareMathOperator{\diag}{diag}

\DeclareMathOperator{\MSE}{MSE}

\usepackage{fancyhdr}
\fancypagestyle{firststyle}{
	\fancyhf{}
	\fancyhead[L]{A. Bazzi and M. Chafii, ``Towards ISAC RIS-Enabled Passive Radar Target Localization'', accepted in \textit{IEEE International Conference on Communications} (ICC), June. 2025.}

}

\def\BibTeX{{\rm B\kern-.05em{\sc i\kern-.025em b}\kern-.08em
    T\kern-.1667em\lower.7ex\hbox{E}\kern-.125emX}}
\begin{document}

\title{Towards ISAC RIS-Enabled Passive Radar Target Localization}

%\author{\IEEEauthorblockN{1\textsuperscript{st} Ahmad Bazzi}
%\IEEEauthorblockA{\textit{Engineering Division} \\
%\textit{New York University Abu Dhabi}\\
%Abu Dhabi, UAE \\
%ahmad.bazzi@nyu.edu}
%\and
%\IEEEauthorblockN{2\textsuperscript{nd} Marwa Chafii}
%\IEEEauthorblockA{\textit{Engineering Division} \\
%\textit{New York University Abu Dhabi}\\
%Abu Dhabi, UAE \\}
%\IEEEauthorblockA{\textit{NYU WIRELESS} \\
%\textit{NYU Tandon School of Engineering}\\
%Brooklyn, NY \\
%marwa.chafii@nyu.edu}
%}
\author{\IEEEauthorblockN{Ahmad Bazzi\IEEEauthorrefmark{1}\IEEEauthorrefmark{2} and Marwa Chafii\IEEEauthorrefmark{1}\IEEEauthorrefmark{2}}
\IEEEauthorblockA{\IEEEauthorrefmark{1}Engineering Division, New York University (NYU), Abu Dhabi, UAE.}
	\IEEEauthorblockA{\IEEEauthorrefmark{2}NYU WIRELESS, NYU Tandon School of Engineering, New York, USA}}

\maketitle
\thispagestyle{firststyle}

\begin{abstract}
Incorporating integrated sensing and communication  capabilities into forthcoming 6G wireless networks is crucial for achieving seamless synchronization between the digital and physical worlds.
The following paper focuses on a scenario where a passive radar (PR) is subject to weak line-of-sight signals of opportunity, emanating from an access point and subsequently reflecting off targets, ultimately reaching the PR.
Furthermore, a normalized least mean squares method is presented for jointly detecting the number of targets and estimating target angles of arrival (AoAs). The algorithm iteratively adjusts the steering vector estimates to minimize a suitable error cost function, while the target AoAs are identified via a peak-finding search conducted on the resulted power spectrum.
Simulation results show the capabilities of the proposed localization method, as well as a $14\dB$ dynamic range reduction that can be achieved at the PR.
\end{abstract}


\section{Introduction}
\label{sec:introduction}
Research on $6$G envisions future services and applications, targeting diverse market demands and disruptive technologies. $6$G aims to support an array of services like blockchain, haptic telemedicine, VR/AR remote services, holographic teleportation, and \ac{XR}. It goes beyond mere connectivity, integrating sensing, computing, and utilizing environmental data for AI and machine learning. Nevertheless, bandwidth-demanding applications necessitate a significant capacity increase, expected to rise by a factor of $10^3$ by $2030$. Enabling technologies like sub-$6$ GHz, \ac{mmWave} frequencies, and \ac{RIS} play crucial roles.

\acp{RIS}, composed of reconfigurable reflecting elements, have garnered attention for their capability to enhance wireless sensor networks' capacity and coverage \cite{renzo2019smart}. These structures manipulate incident signals in the wireless propagation environment, altering wireless channels between network nodes. \acp{RIS} offer low energy consumption, allowing signal amplification without power amplifiers. Yet, their low cost might lead to large-scale deployment, sprawling entire building walls \cite{han2022localization}. Properly designed phase shifts from each reflecting element combine reflected signals effectively. They are known as \ac{LISA} or \ac{IRS} in literature. RISs have diverse applications in wireless communications, such as beamforming, security, OFDM, and millimeter-wave channel estimation \cite{lin2020reconfigurable, xu2022ris, zhou2022channel}. Their deployment enhances spectral and energy efficiency, aids in environment-friendly deployment, and offers secure wireless communication \cite{huang2019reconfigurable,liaskos2018using}. Over the last 80 years, passive radar, which localizes targets without emitting controlled radar data, has been extensively studied \cite{dawidowicz2012detection,strinati2021wireless}. RIS is anticipated to be pivotal in advanced localization and sensing \cite{10061453} applications via passive radar. Passive radar offers cost-effective procurement, operation, and maintenance, as well as covert operation. However, it faces challenges in target localization due to the absence of transmit waveform information. Nonetheless, it is protocol-agnostic and lacks active component assistance in some instances.
%% Existing work

Channel estimation, typically between any two nodes within the system, is essential for designing the reflective matrix, which involves phase shifters at the \ac{RIS}. One approach, as in \cite{taha2021enabling}, employs deep learning and compressive sensing techniques for \ac{RIS} phase shift design without requiring channel estimation for target localization. An alternative approach in \cite{he2020adaptive} suggests an adaptive RIS phase shifter using hierarchical codebooks with UE feedback for localization, yet it differs fundamentally from our scenario. Specifically, it doesn't involve \ac{PR} or leverage multipath bounces off targets for localization. Similarly, works such as \cite{fascista2021ris} and \cite{elzanaty2021reconfigurable} perform localization with RIS but without \ac{PR}, where the latter also lacks reflections from the targets. In contrast to these approaches, our proposed model involves \ac{PR} performing spatial beamforming towards the \ac{RIS}, focusing on reflections bouncing off the targets towards the \ac{RIS}, which redirects these bounces towards the \ac{PR}. Conversely, works such as \cite{zhang2021metalocalization} optimize \ac{RIS} phase shift coefficients using \ac{RSS} values to maximize location change differences. 

%% This paper
The focus of this paper is on a situation where a \ac{PR} is exposed to weak \ac{LoS} signals of opportunity created by communication signals transmitted by an \ac{AP}, and further reflected off targets, ultimately reaching the \ac{PR}.
Indeed, \ac{PR} can suffer from short operational range and low range resolution problems \cite{9625027}, in addition to distant targets, which tend to have obstructed or very weak \ac{LoS} signal \cite{10096466}.
To address this, a \ac{RIS} is deployed to cover the entire environment, establishing robust \ac{LoS} connections with both the targets and the \ac{PR}. To this end, the contributions of the paper can be summarized as follows: 
We formulate an appropriate system model to represent the aforementioned situation.
Due to the power discrepancies within the received signal, the \ac{RIS} modifies its phase shifts in coordination with \ac{PR} beamforming, ensuring the robust reception of the \ac{LoS} signals for further localization, while minimizing the required \ac{DR} at the \ac{PR}.
Furthermore, an adaptive approach based on \ac{NLMS} is introduced, aiming to concurrently discern the number of targets and estimate their respective \acp{AoA}. Through iterative refinement of the steering vector estimates, the algorithm minimizes a suitable error cost function, while the target \acp{AoA} are identified via a peak-finding search conducted on the resulted power spectrum.
The simulation results demonstrate the efficacy of the proposed localization method and reveal a notable $14\dB$ reduction in \ac{DR} achievable at the \ac{PR}.
%% TODO: Continue if any

%% TODO: Organization
The following paper is organized as follows: 
Section \ref{sec:system-model} introduces the system model of the underlying problem.
Section \ref{sec:ris-pr-optimization} optimizes the \ac{RIS} phase shifts as well as the \ac{PR} beamforming weights to achieve a low \ac{DR} at the \ac{PR}.
Section \ref{sec:localization}, we describe a technique for simultaneously detecting the number of targets and estimating target \acp{AoA}.
Section \ref{sec:simulations} presents our simulation results.

\textbf{Notation}: Upper-case and lower-case boldface letters denote matrices and vectors, respectively. $(.)^T$, $(.)^*$ and $(.)^H$ represent the transpose, the conjugate and the transpose-conjugate operators. The statistical expectation is denoted as $\mathbb{E}\lbrace \rbrace$. For any complex number $z \in \mathbb{C}$, the magnitude is denoted as $\vert z \vert$, its angle is $ \angle z$. The $\ell_2$ norm of a vector $\pmb{x}$ is denoted as $\Vert \pmb{x} \Vert$. The matrix $\pmb{I}_N$ is the identity matrix of size $N \times N$. The zero-vector is $\pmb{0}$. For matrix indexing, the $(i,j)^{th}$ entry of matrix $\pmb{A}$ is denoted by $\pmb{A}_{i,j}$. The sub-matrix of $\pmb{A}$ ranging from rows $i_1$ to $i_2>i_1$ and columns $j_1$ to $j_2>j_1$ is denoted as $\pmb{A}_{i_1:i_2,j_1:j_2}$.

\section{System Model}
\label{sec:system-model}
\begin{figure}[!t]
\centering
\includegraphics[width=3.5in]{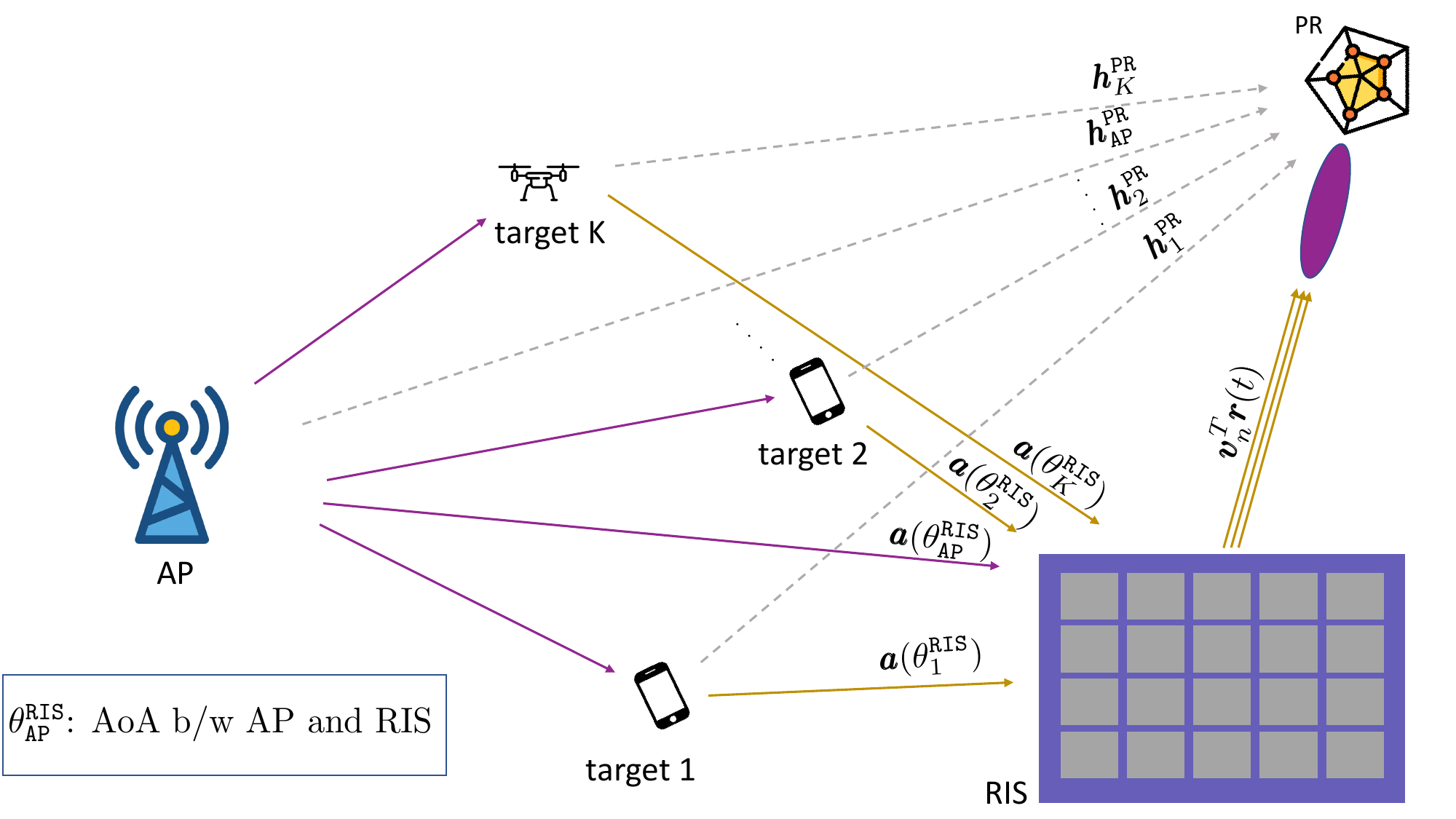}
\caption{The scenario involves an \ac{RIS}-based multi-target system, where an \ac{AP} serves communication users in the downlink. The system includes an \ac{RIS} with $M$ elements, which reflects signals bouncing off targets towards the \ac{PR}. 
\textbf{The dashed lines model weak \ac{LoS} components, whereas the solid lines model a strong \ac{LoS} channel.}}
\label{fig_1}
\end{figure}

Let's consider a scenario involving a narrow-band system tasked with localizing $K$ targets. In this context, communication users are also considered targets. 
To simplify, consider a single-antenna \ac{AP} transmitting a signal, denoted as $s(t)$. 
In an \ac{ISAC} framework, the signal $s(t)$ initially intended for communication purposes, is also exploited by the radar sub-systems \cite{10417003}, specifically the \ac{PR}. 
In the \ac{DL} transmission, the \ac{AP} broadcasts $s(t)$ through an omni-directional antenna. 
Consequently, the signal reaches and bounces off the $K$ targets, whereby the \ac{RIS} receives those reflections, along with the direct signal path between the \ac{AP} and \ac{RIS} itself. 
As a result, the signal received at the RIS can be described as
\begin{equation}
\label{eq:rt}
	\pmb{r}(t) 
	= 
	\pmb{A}(\Theta_{1:K}^{\tt{RIS}})
	\pmb{s}(t)
	+
	\alpha_0
	\pmb{a}_{M}(\theta_{\tt{AP}}^{\tt{RIS}})
	s(t - \tau_{\tt{AP}}^{\tt{RIS}}),
\end{equation}
where the \ac{LoS} element is denoted by $\pmb{a}_M(\theta_k^{\tt{RIS}})$, where $\theta_k^{\tt{RIS}}$ signifies the angle between the \ac{RIS} and the $k^{th}$ target.
In fact, the complex vector $\pmb{a}_M(\theta) \in \mathbb{C}^{M \times 1}$ denotes the input array steering vector for the \ac{RIS} toward an \ac{AoA} $\theta$, which is represented as
\begin{equation}
	\label{eq:Rx-Steering-Vector}
	\pmb{a}_M(\theta)
	=
	\begin{bmatrix}
		a_1(\theta) & \ldots & a_M(\theta)
	\end{bmatrix}^T.
\end{equation}
Likewise, $\pmb{a}_{M}(\theta_{\tt{AP}}^{\tt{RIS}})$ is the \ac{LoS} component between AP and \ac{RIS}, $\theta_{\tt{AP}}^{\tt{RIS}}$ is the \ac{AoA} between \ac{AP} and \ac{RIS}, and $\bar{\pmb{h}}_{\tt{AP}}^{\tt{RIS}} $ is the non-\ac{LoS} component between AP and \ac{RIS}. In addition, $\kappa_{\tt{AP}}^{\tt{RIS}}$ is the Rician factor for the channel between AP and \ac{RIS}.
%For convenience, we express $\pmb{H}_{\tt{tar}}^{\tt{RIS}}$ as 
%\begin{equation}
%	\label{eq:equivalent-channel}
%	\pmb{H}_{\tt{tar}}^{\tt{RIS}}
%	=
%	\pmb{A}(\Theta_{1:K}^{\tt{RIS}}) 
%	\pmb{\kappa}
%	+
%	\bar{\pmb{H}}_{\tt{tar}}^{\tt{RIS}}
%	\bar{\pmb{\kappa}},
%\end{equation}
%where
%\begin{align}
%	\pmb{\kappa} &= \diag \begin{bmatrix}\sqrt{\frac{\kappa_1}{\kappa_1 + 1}} & \ldots & \sqrt{\frac{\kappa_K}{\kappa_K + 1}} \end{bmatrix}^T, \\
%	\bar{\pmb{\kappa}} &= \diag \begin{bmatrix}\sqrt{\frac{1}{\kappa_1 + 1}} & \ldots & \sqrt{\frac{1}{\kappa_K + 1}} \end{bmatrix}^T.
%\end{align}
In addition, $\pmb{A}(\Theta_{1:K}^{RIS})$ is the steering matrix, deriving from the combined contribution of each path between the targets and the \ac{RIS}, which is given as $\pmb{A}(\Theta_{1:K}^{\tt{RIS}}) 
	=
	\begin{bmatrix}
		\pmb{a}_{M}(\theta_{1}^{\tt{RIS}}) 
		& \pmb{a}_{M}(\theta_{2}^{\tt{RIS}}) 
		& 
		\ldots 
		&
		\pmb{a}_{M}(\theta_{K}^{\tt{RIS}})
	\end{bmatrix}$. 
Here, the steering vector $\pmb{a}_{M}(\theta) \in \mathbb{C}^{M \times 1}$, characterizes the steering at the output of the $M$ elements within the \ac{RIS}.
The signal $\pmb{s}(t) \in \mathbb{C}^{K \times 1}$ contains scaled and delayed versions of the original transmitted signal $s(t)$, where its $k^{th}$ entry is given as $[\pmb{s}(t)]_k = \alpha_k s(t - \tau_{\tt{AP}}^{k} -\tau_{k}^{\tt{RIS}})$,
where the following parameters encapsulate the propagation delays within the system, i.e. $\tau_{\tt{AP}}^{k}$ is the delay between the \ac{AP} and the $k^{th}$ target, $\tau_{k}^{\tt{RIS}}$ is the delay between the $k^{th}$ target and the \ac{RIS}, and $\tau_{\tt{AP}}^{\tt{RIS}}$ is the delay the \ac{AP} and the \ac{RIS}. 
Moreover, $\alpha_k$ represents the reflection gain, encompassing large-scale fading effects of the two-way \ac{AP}-target-\ac{RIS}. However, $\alpha_0$ serves as a complex coefficient considering the direct channel gain specifically between the \ac{AP} and the \ac{RIS}.
To this end, the expression describing the signal reflected by the \ac{RIS} towards the \ac{PR} is presented as
\begin{equation}
\label{eq:RIS-model-with-Theta}
	x_n(t) 
	=
\pmb{b}_{M}^T(\phi_{\tt{RIS}}^{\tt{PR}})
	\pmb{\Theta}^{(n)} 
	\Big(
	\pmb{A}(\Theta_{1:K}^{\tt{RIS}})
	\pmb{s}(t)
	+
	\alpha_0
	\pmb{a}_{M}(\theta_{\tt{AP}}^{\tt{RIS}})
	s(t - \tau_{\tt{AP}}^{\tt{RIS}})
	 \Big).
\end{equation}
Here, $\pmb{\Theta}^{(n)} = \text{diag}(e^{j\beta_1^{(n)}}, e^{j\beta_2^{(n)}} \ldots e^{j\beta_M^{(n)}})$ represents the \ac{RIS} shifting elements during the $n^{th}$ epoch. An epoch signifies a time period where a realization of $\pmb{r}(t)$, as defined in equation \eqref{eq:rt}, is received.
Therefore, $\beta_m^{(n)}$ specifically denotes the phase shift at the $m^{th}$ \ac{RIS} element during the epoch $n$.
The vector $\pmb{b}_{M}(\phi) \in \mathbb{C}^{M \times 1}$ defines the output transmit steering vector of the \ac{RIS}, articulated as $\pmb{b}_{M}(\phi) = \begin{bmatrix} b_1(\phi) & \ldots & b_M(\phi) \end{bmatrix}$.
The angle $\phi_{\tt{RIS}}^{\tt{PR}}$ represents the \ac{AoD} from the \ac{RIS} to the \ac{PR}. Particularly, in the scenario assuming a \ac{LoS} channel between the \ac{RIS} and the \ac{PR}, the channel between each \ac{RIS} element and the \ac{PR} is denoted by $\pmb{b}_{M}^T(\phi_{\tt{RIS}}^{\tt{PR}})$.
Finally, the signal received at the \ac{PR} can be represented as
\begin{equation}
\begin{split}
	\label{eq:rx-signal-PR}
	\pmb{y}_n(t)
	&=
	\rho_{\tt{RIS}}^{\tt{PR}}\pmb{a}(\theta_{\tt{RIS}}^{\tt{PR}})x_n(t - \tau_{\tt{RIS}}^{\tt{PR}}) 
	+
	\rho_{\tt{AP}}^{\tt{PR}} 
	\pmb{h}_{\tt{AP}}^{\tt{PR}}
	s(t-\tau_{\tt{AP}}^{\tt{PR}}) \\
	&+
	\sum\nolimits_{k=1}^K
	\rho_k
	\pmb{h}_{k}^{\tt{PR}}
	s(t-\tau_{\tt{AP}}^{k}-\tau_k^{\tt{PR}})
	+\pmb{\epsilon}_n(t).
\end{split}
\end{equation}
The channel $\pmb{h}_{\tt{AP}}^{\tt{PR}}$ is described as a combination of the \ac{LoS} component, denoted by $\sqrt{\frac{\kappa_{\tt{AP}}^{\tt{PR}}}{1 + \kappa_{\tt{AP}}^{\tt{PR}} }}\pmb{a}(\theta_{\tt{AP}}^{\tt{PR}})$, and the non-\ac{LoS} component $\sqrt{\frac{1}{1 + \kappa_{\tt{AP}}^{\tt{PR}} }}\bar{\pmb{h}}_{\tt{AP}}^{\tt{PR}}$, namely 
\begin{equation}
	\pmb{h}_{\tt{AP}}^{\tt{PR}}
	=
	\sqrt{\frac{\kappa_{\tt{AP}}^{\tt{PR}}}{1 + \kappa_{\tt{AP}}^{\tt{PR}} }}\pmb{a}(\theta_{\tt{AP}}^{\tt{PR}})
	+
	\sqrt{\frac{1}{1 + \kappa_{\tt{AP}}^{\tt{PR}} }}\bar{\pmb{h}}_{\tt{AP}}^{\tt{PR}}.
\end{equation}
Here, $\kappa_{\tt{AP}}^{\tt{PR}}$ characterizes the Rician factor representing the channel between the \ac{AP} and the \ac{PR}, while $\bar{\pmb{h}}_{\tt{AP}}^{\tt{PR}}$ refers to the non-\ac{LoS} part of that channel.
Similarly, the expression for $\pmb{h}_{k}^{\tt{PR}}$ encompasses the Rician factor $\kappa_{k}^{\tt{PR}}$ for the $k^{th}$ target and the \ac{PR}. 
It involves the \ac{LoS} component $\sqrt{\frac{\kappa_{k}^{\tt{PR}}}{1 + \kappa_{k}^{\tt{PR}} }}\pmb{a}(\theta_{k}^{\tt{PR}})$ and the non-\ac{LoS} component $\sqrt{\frac{1}{1 + \kappa_{k}^{\tt{PR}} }}\bar{\pmb{h}}_{k}^{\tt{PR}}$, representing the non-\ac{LoS} link between the $k^{th}$ target and the \ac{PR}, i.e.
\begin{equation}
	\pmb{h}_{k}^{\tt{PR}}
	=
	\sqrt{\frac{\kappa_{k}^{\tt{PR}}}{1 + \kappa_{k}^{\tt{PR}} }}\pmb{a}(\theta_{k}^{\tt{PR}})
	+
	\sqrt{\frac{1}{1 + \kappa_{k}^{\tt{PR}} }}\bar{\pmb{h}}_{k}^{\tt{PR}}.
\end{equation}
In equation \eqref{eq:rx-signal-PR}, the vector $\pmb{y}_n(t) \in \mathbb{C}^{N_{PR} \times 1}$ signifies the received signal at the \ac{PR}.
Summarizing the system model as illustrated in Fig. \ref{fig_1}, the terms within the equation represent different aspects, which are the following; the first term, $\rho_{\tt{RIS}}^{\tt{PR}}\pmb{a}(\theta_{\tt{RIS}}^{\tt{PR}})x_n(t - \tau_{\tt{RIS}}^{\tt{PR}})$, capsulizes all the sensing data collected at the \ac{RIS} and reflected back to the \ac{PR}. 
The second term, $\rho_{\tt{AP}}^{\tt{PR}}\pmb{h}_{\tt{AP}}^{\tt{PR}}s(t-\tau_{\tt{AP}}^{\tt{PR}})$, represents the channel between the AP and the \ac{PR}. The third term, $\sum_{k=1}^K\rho_k\pmb{h}_{k}^{\tt{PR}}s(t-\tau_{\tt{AP}}^{k}-\tau_k^{\tt{PR}})$, involves the contributions from the $K$ targets. 
These signals propagate from the \ac{AP} to the targets and then bounce back towards the \ac{PR}. 
Here, $\tau_{\tt{RIS}}^{\tt{PR}}$ refers to the delay between the \ac{RIS} and the \ac{PR}, $\tau_{\tt{AP}}^{\tt{PR}}$ is the delay between the AP and the \ac{PR}, and $\tau_k^{\tt{PR}}$ signifies the delay between the $k^{th}$ target and the \ac{PR}.
Moreover, the large-scale fading gains, represented by $\rho_k$, are defined in a similar manner as the $\alpha_k$s. 
Furthermore, the term $\pmb{\epsilon}_n(t)$ represents the noise, particularly \ac{AWGN}. Sampling equation \eqref{eq:rx-signal-PR} at $L$ time instances and concatenating the samples provides
\begin{equation}
\label{eq:Rx-data-at-PR}
\begin{split}
	\pmb{Y}_n
	&=
	\begin{bmatrix}
		\pmb{y}_n(t_1) & \pmb{y}_n(t_2) & \ldots & \pmb{y}_n(t_L)
	\end{bmatrix}  % Y{n} in MATLAB	
	 \\
	&= 
	\rho_{\tt{RIS}}^{\tt{PR}}
	\pmb{a}(\theta_{\tt{RIS}}^{\tt{PR}})
	\pmb{x}_n
	+
	\rho_{\tt{AP}}^{\tt{PR}} 
	\pmb{h}_{\tt{AP}}^{\tt{PR}}
	\pmb{s}_{\tt{AP}}^{\tt{PR}}
	+
	\sum\nolimits_{k=1}^K
	\rho_k
	\pmb{h}_{k}^{\tt{PR}}
	\pmb{s}_k^{\tt{PR}}
	+
	\pmb{E}_n.
\end{split}
\end{equation}
Equation \eqref{eq:Rx-data-at-PR} defines the matrix $\pmb{Y}_n \in \mathbb{C}^{N_{\tt{PR}} \times L}$ as the collected sampled data matrix at the \ac{PR} for the epoch $n$. In addition, the terms appearing in \eqref{eq:Rx-data-at-PR} are defined as 
\begin{align}
	\pmb{x}_n &= 
	\begin{bmatrix}
		x_n(t_1) & x_n(t_2) & \ldots & x_n(t_L)
	\end{bmatrix}, \\
	\pmb{s}_{\tt{AP}}^{\tt{PR}} &= 
	\begin{bmatrix}
		s(t_1-\tau_{\tt{AP}}^{\tt{PR}}) &
		s(t_2-\tau_{\tt{AP}}^{\tt{PR}}) &
		\ldots &
		s(t_L-\tau_{\tt{AP}}^{\tt{PR}})
	\end{bmatrix} ,\\
	\pmb{s}_k^{\tt{PR}} 
	&=
	\begin{bmatrix}
		s(t_1-\tau_{\tt{AP}}^{k}-\tau_k^{\tt{PR}})  \\
		s(t_2-\tau_{\tt{AP}}^{k}-\tau_k^{\tt{PR}})  \\
		\vdots \\
		s(t_L-\tau_{\tt{AP}}^{k}-\tau_k^{\tt{PR}}) 
	\end{bmatrix}^T ,\\
	\pmb{E}_n &= 
	\begin{bmatrix}
		\pmb{\epsilon}_n(t_1) & \pmb{\epsilon}_n(t_2) & \ldots & \pmb{\epsilon}_n(t_L)
	\end{bmatrix}^T .
\end{align}
Collecting all samples at different epochs in one data matrix, we now have the following at the \ac{PR},
\begin{equation}
	\pmb{Y} 
	=
	\begin{bmatrix}
		\pmb{Y}_1^T &
		\pmb{Y}_2^T &
		\ldots &
		\pmb{Y}_{N_{\text{epoch}}}^T
	\end{bmatrix}^T
	\in
	\mathbb{C}^{N_{\text{PR}}N_{\text{epoch}} \times L}.
\end{equation}
\section{\ac{RIS} and \ac{PR} Optimization}
\label{sec:ris-pr-optimization}
In this section, we describe how the \ac{RIS} the phase shifts to re-direct useful signals for localization towards the \ac{PR}. In turn, we also describe how the \ac{PR} can optimize its analog beamformer to receive the \ac{RIS} signals with minimal interference. For the \ac{RIS}, we propose the following
\begin{equation}
 \label{eq:CancelAP-RIS}
\begin{aligned}
(\mathcal{P}_{\tt{RIS}}):
\begin{cases}
\min\limits_{ \beta_m^{(n)} }&  \Vert \pmb{b}_M^T(\phi_{\tt{RIS}}^{\tt{PR}}) \pmb{\Theta}^{(n)} 
\pmb{a}_{M}(\theta_{\tt{AP}}^{\tt{RIS}}) \Vert^2   \\
\textrm{s.t.}
 & \beta_m^{(n)} \in [-\pi,\pi], \quad \forall \ m
\end{cases}
\end{aligned}
\end{equation}
The RIS phase-shifter matrix is employed to \textit{minimize the known contribution, i.e. AP-RIS-PR (experiencing double path-loss), which is a fixed path with considerably higher power compared to other paths, such as the AP-target-RIS-PR (experiencing triple path-loss) path.} This entails that an unnecessarily high \ac{DR} would be needed at the \ac{PR} to discern the smaller paths arising from AP-target-RIS-PR.
The optimization problem presented in \eqref{eq:CancelAP-RIS} aims to reduce the channel contribution between the AP and the \ac{RIS}, rather than forming beams towards specific regions. The equivalent optimization problem $(\mathcal{P}_{\tt{RIS}})$ can be expressed as follows
\begin{equation}
 \label{eq:CancelAP-RIS}
\begin{aligned}
(\mathcal{P}_{\tt{RIS}}):
\begin{cases}
\min\limits_{ \pmb{v}_n }&  \Vert \pmb{v}_n^T 
\widetilde{\pmb{a}}_{M}(\theta_{\tt{AP}}^{\tt{RIS}}) \Vert^2   \\
\textrm{s.t.}
 & \vert [\pmb{v}_n]_m \vert = 1, \quad \forall m,
\end{cases}
\end{aligned}
\end{equation}
where $\widetilde{\pmb{a}}_{M}(\theta_{\tt{AP}}^{\tt{RIS}}) = \pmb{B} {\pmb{a}}_{M}(\theta_{\tt{AP}}^{\tt{RIS}})$ and $\pmb{B} = \diag[ \pmb{b}_M(\phi_{\tt{RIS}}^{\tt{PR}}) ]$. Note that $[\pmb{v}_n]_m = e^{j \beta_m^{(n)}}$, thanks to the constant modulus constraint. 
In addition, this optimization problem acts on an instantaneous basis, i.e. it computes the \ac{RIS} phase shift coefficients, i.e. $\beta_m^{(n)}$ (equivalently $[\pmb{v}_n]_m$) at epoch $n$. To optimize over the available epochs, the following optimization problem is proposed
\begin{equation}
 \label{eq:CancelAP-RIS-Batch}
\begin{aligned}
(\mathcal{P}_{\tt{RIS}}):
\begin{cases}
\min\limits_{ \pmb{V} }&  \Vert \pmb{V} 
\widetilde{\pmb{a}}_{M}(\theta_{\tt{AP}}^{\tt{RIS}}) \Vert^2    \\
\textrm{s.t.}
 & \vert \pmb{V}_{n,m} \vert  = 1, \quad \forall n,m.
\end{cases}
\end{aligned}
\end{equation}
In equation \eqref{eq:CancelAP-RIS-Batch}, the matrix $\pmb{V} = \begin{bmatrix} \pmb{v}_1  & \pmb{v}_2&   \ldots & \pmb{v}_{N_{\tt{epoch}}}\end{bmatrix}^T \in \mathbb{C}^{N_{\tt{epoch}} \times M}$ concatenates all the phase-shift vectors. To this end, the optimization problem $(\mathcal{P}_{\tt{RIS}})$ is non-convex due to the constant-modulus constraint. For this purpose, we suggest to: $1)$ solve the unconstrained optimization problem, then $2)$ adapt the solution towards a feasible one. Towards this goal, the unconstrained version of problem $(\mathcal{P}_{\tt{RIS}})$ admits as solution, 
\begin{equation}
	\overline{\pmb{V}} = \mathcal{P}^{\perp}_{\widetilde{\pmb{a}}_{M}(\theta_{\tt{AP}}^{\tt{RIS}})} = \pmb{I}_M -
	\big\Vert \widetilde{\pmb{a}}_{M}(\theta_{\tt{AP}}^{\tt{RIS}}) \Vert^2  \big\Vert^{-2}  \widetilde{\pmb{a}}_{M}(\theta_{\tt{AP}}^{\tt{RIS}})
	\widetilde{\pmb{a}}_{M}^H(\theta_{\tt{AP}}^{\tt{RIS}}) 
	.
\end{equation}
The above matrix requires a one-time computation. Furthermore, the solution is in-feasible, as there is no warranty on $\overline{\pmb{V}}_{i,j}$ to satisfy the constant modulus constraint. In order to project onto the feasible set, the following solution is proposed
\begin{equation}
\label{eq:V-solution}
	\pmb{V} = 
	\exp
	\Big\lbrace
	j \angle (\mathcal{P}^{\perp}_{\widetilde{\pmb{a}}_{M}(\theta_{\tt{AP}}^{\tt{RIS}})}  \pmb{\Gamma} )
	\Big\rbrace,
\end{equation}
where $\pmb{\Gamma} \in \mathbb{C}^{M \times N_{\text{epoch}}}$ is a random matrix drawn from a standard Gaussian distribution.
From equation \eqref{eq:V-solution}, we obtain $\beta_m^{(n)}$ as 
\begin{equation}
	\label{eq:beta-optimal-vals}
	\beta_m^{(n)}
	=
	\Big[ \angle (\mathcal{P}^{\perp}_{\widetilde{\pmb{a}}_{M}(\theta_{\tt{AP}}^{\tt{RIS}})}  \pmb{\Gamma} )
	\Big]_{n,m}, \quad \forall n,m.
\end{equation}
In the above, note that $\beta_m^{(n)}$ is independent of the signal structure of $s(t)$. This suggests that no apriori knowledge is required at the \ac{PR} and \ac{RIS}.
Besides the \ac{RIS}, the \ac{PR} beamformers towards the \ac{RIS} to listen to all AP-target-RIS-PR paths. More specifically, the \ac{PR} applies a static beamforming vector $\pmb{w}$ as follows
\begin{equation}
\begin{split}
	\pmb{w}^H\pmb{Y}_n
	&=
	\rho_{\tt{RIS}}^{\tt{PR}}
\pmb{w}^H
	\pmb{a}(\theta_{\tt{RIS}}^{\tt{PR}})
	\pmb{x}_n \\
	&+
\pmb{w}^H
\Big(
	\rho_{\tt{AP}}^{\tt{PR}} 
	\pmb{h}_{\tt{AP}}^{\tt{PR}}
	\pmb{s}_{\tt{AP}}^{\tt{PR}}
	+
	\sum\nolimits_{k=1}^K
	\rho_k
	\pmb{h}_{k}^{\tt{PR}}
	\pmb{s}_k^{\tt{PR}}
	+
	\pmb{E}_n
	\Big).
\end{split}
\end{equation}
over all epochs $n = 1 \ldots N_{\text{epoch}}$. The first term is the signal-of-interest needed at the \ac{PR} to localize the targets, whereas the second and third terms contain weak \ac{LoS} signals that can not be reliably used for localization.
Although a simple choice of $\pmb{w}
	=
	\frac{\pmb{a}(\theta_{\tt{RIS}}^{\tt{PR}})}{\Vert \pmb{a}(\theta_{\tt{RIS}}^{\tt{PR}}) \Vert^2 }$ is straightforward, a significant illumination towards the \ac{RIS} is attained by spatially filtering out other contributions not reflected by the \ac{RIS}.
After beamforming at each time epoch, the collected data are stacked as
\begin{equation}
\begin{split}
	\pmb{Z}
	&=
	\begin{bmatrix}
		\pmb{Y}_1^H\pmb{w}  &
		\pmb{Y}_2^H\pmb{w}  &
		\ldots &
		\pmb{Y}_{N_{\tt{epoch}}}^H\pmb{w} 
	\end{bmatrix}^H\\
	&=
	\begin{bmatrix}
		\pmb{z}_1 & \pmb{z}_2 & \ldots & \pmb{z}_L
	\end{bmatrix},
\end{split}
\end{equation}
where $\pmb{Z} \in \mathbb{C}^{N_{\tt{epoch}} \times L}$ is the beamformed data matrix at the \ac{PR} over all epochs and time samples. 
In the next section, we introduce a method for joint detection of the number of targets and target \ac{AoA} estimation. The algorithm iteratively updates the steering vector to minimize the error cost function, and a peak-finding search on the computed power spectrum yields estimates for the target \acp{AoA}.

\section{Target Detection and Localization}
\label{sec:localization}
In this section, we describe a \ac{NLMS} approach towards simulateneous detection of number of targets and target localization. 
First, given a matrix $\pmb{Z}$ representing the data received at the \ac{PR}, the objective is to estimate all angles within $\pmb{\Theta}_{1:K}^{\tt{RIS}}$, encompassing all target Angle of Arrival (\ac{AoA}) information. However, in various scenarios, particularly dynamic environments, the number of targets denoted as $K$ may be unknown. In this context, consider the output of the sum-and-delay beamformer at a specific direction $\theta$, utilizing the steering vector $\pmb{V}\pmb{a}_M(\theta)$ at the $\ell^{th}$ snapshot, expressed as
\begin{equation}
	\label{eq:pelltheta}
	p_{\ell}(\theta) 
	=
	\pmb{a}_M^H(\theta)
	\pmb{V}^H
	\pmb{z}_\ell.
\end{equation}
Subsequently, we can formulate the error cost function that we aim to minimize in each snapshot, namely $C(\ell) = \mathbb{E}( \vert e(\ell) \vert^2 )$, where $e(\ell) = p_{\ell}(\theta) - \pmb{a}^H \pmb{z}_\ell$. The \ac{LMS} aims at minimizing the above cost. First, we take the gradient with respect to $\pmb{a}$ to get 
\begin{equation*}
	\begin{split}
		\nabla_{\pmb{a}} C(\ell) &= 2 \mathbb{E} \lbrace \nabla_{\pmb{a}} [ e(\ell) ]e^*(\ell) \rbrace= -2 \mathbb{E} \lbrace ( p_{\ell}(\theta) - \pmb{a}^H \pmb{z}_\ell )^* \pmb{z}_\ell \rbrace,
	\end{split}
\end{equation*}
where the last step is due to the gradient's value, $\nabla_{\pmb{a}} [ e(\ell) ] = -\pmb{z}_\ell$. Thanks to $\nabla_{\pmb{a}} C(\ell) $, the \ac{LMS} filter can now focus towards the steepest ascent of $C(\ell)$ by taking a direction opposite to $\nabla_{\pmb{a}} C(\ell)$, namely
\begin{equation}
\begin{split}
	\hat{\pmb{a}}_{\ell + 1}(\theta)
	&=
	\hat{\pmb{a}}_{\ell}(\theta)
	-
	\frac{\mu}{2} \big[ \nabla_{\pmb{a}} C(\ell) \big]_{\pmb{a} = \hat{\pmb{a}}_{\ell}} \\
	&= \hat{\pmb{a}}_{\ell}(\theta)
	+
	\mu \mathbb{E} \lbrace ( p_{\ell}(\theta) - \hat{\pmb{a}}_{\ell}^H(\theta) \pmb{z}_\ell )^* \pmb{z}_\ell \rbrace,
\end{split}
\end{equation}
where $\frac{\mu}{2}$ is the well-known \ac{LMS} adaptation constant, i.e. the step size. Now that we have an update equation for the steering vector in the $\theta$'s look-direction, the only problem that remains is the expectation operator, which is not available at time update $\ell$. Therefore, omitting it we finally have
\begin{equation}
	\hat{\pmb{a}}_{\ell + 1}(\theta)
	=
	\hat{\pmb{a}}_{\ell}(\theta)
	+
	\mu ( p_{\ell}(\theta) - \hat{\pmb{a}}_{\ell}^H(\theta) \pmb{z}_\ell )^* \pmb{z}_\ell .
\end{equation}
\begin{algorithm}
\caption{Batch NLMS filter Localization}\label{alg:cap}
\begin{algorithmic}[1]
\State \textbf{INPUT} $\pmb{Z} = \begin{bmatrix} \pmb{z}_1 & \ldots & \pmb{z}_L \end{bmatrix}$, $\mu$, $\Theta_{\tt{grid}}$
\While{$\theta \neq \emptyset$}
\State $\hat{\pmb{a}}_{0}(\theta) \gets \pmb{0}_{N_{\tt{epoch}} \times 1}$  
\State $\ell \gets 0$
\While{$\ell < L$}
\State $p_{\ell}(\theta) 
	\gets
	\pmb{a}_M^H(\theta)
	\pmb{V}^H
	\pmb{z}_\ell$
\State $  \hat{\pmb{a}}_{\ell + 1}(\theta)
	\gets
	\hat{\pmb{a}}_{\ell}(\theta)
	+
	\frac{\mu}{\Vert \pmb{z}_{\ell} \Vert} ( p_{\ell}(\theta) - \hat{\pmb{a}}_{\ell}^H(\theta) \pmb{z}_\ell )^* \pmb{z}_\ell $
\State $\ell \gets \ell + 1$
\EndWhile
\State$P(\theta) = \Vert \hat{\pmb{a}}_{L}(\theta) \Vert^2 $
\State Choose next $\theta \in \Theta_{\tt{grid}}$
\EndWhile
\If{$P(\theta)$ is peak}
\State Insert $\theta$ in $\hat{\pmb{\Theta}}_{1:K}^{\tt{RIS}}$
\EndIf
\State 
\Return $\hat{\pmb{\Theta}}_{1:K}^{\tt{RIS}}$
\end{algorithmic}
\end{algorithm}
A normalized \ac{LMS}, i.e. \ac{NLMS}, could also be proposed at this point to avoid sensitivities of the input norm, i.e. $\Vert \pmb{z}_\ell \Vert$, which makes it extremely difficult to choose an adaptation constant for algorithm stability purposes \cite{haykin1996adaptive}. To this end, the following update is proposed
\begin{equation}
	\hat{\pmb{a}}_{\ell + 1}(\theta)
	=
	\hat{\pmb{a}}_{\ell}(\theta)
	+
	\frac{\mu}{\Vert \pmb{z}_{\ell} \Vert} ( p_{\ell}(\theta) - \hat{\pmb{a}}_{\ell}^H(\theta) \pmb{z}_\ell )^* \pmb{z}_\ell .
\end{equation}
As $\hat{\pmb{a}}(\theta)$ encompasses signal contributions at direction $\theta$, a meaningful metric would be to quantify the overall power within the estimate $\hat{\pmb{a}}_{L}(\theta)$, denoted as
\begin{equation}
	P(\theta) = \Vert \hat{\pmb{a}}_{L}(\theta) \Vert^2 .
\end{equation}
Ultimately, the computation of the $K$ target \acp{AoA} is performed by conducting a one-dimensional peak-finding search on $P(\theta)$, namely
\begin{equation}
	\hat{\pmb{\Theta}}_{1:K}^{\tt{RIS}}
	=
	\arg\max_{\theta_1 \ldots \theta_K} P(\theta).
\end{equation}
A summary of an implementation of the proposed \ac{LMS} filtering approach is summarized in {\tt{\textbf{Algorithm \ref{alg:cap}}}}. One interesting feature of this batch approach is that it could be parallelized over angles to be evaluated on a grid, i.e. $\Theta_{\tt{grid}}$.
In the process of peak detection, it is essential to normalize the entire spectrum $P(\theta)$ to ensure its maximum value is set to $1$. Then, peak identification is done by selecting peaks surpassing a specified threshold value, denoted as $0 < \varphi < 1$.

\section{Simulation Results}
\label{sec:simulations}
\begin{figure}[t!]
	\centering
	\includegraphics[width=0.8\linewidth]{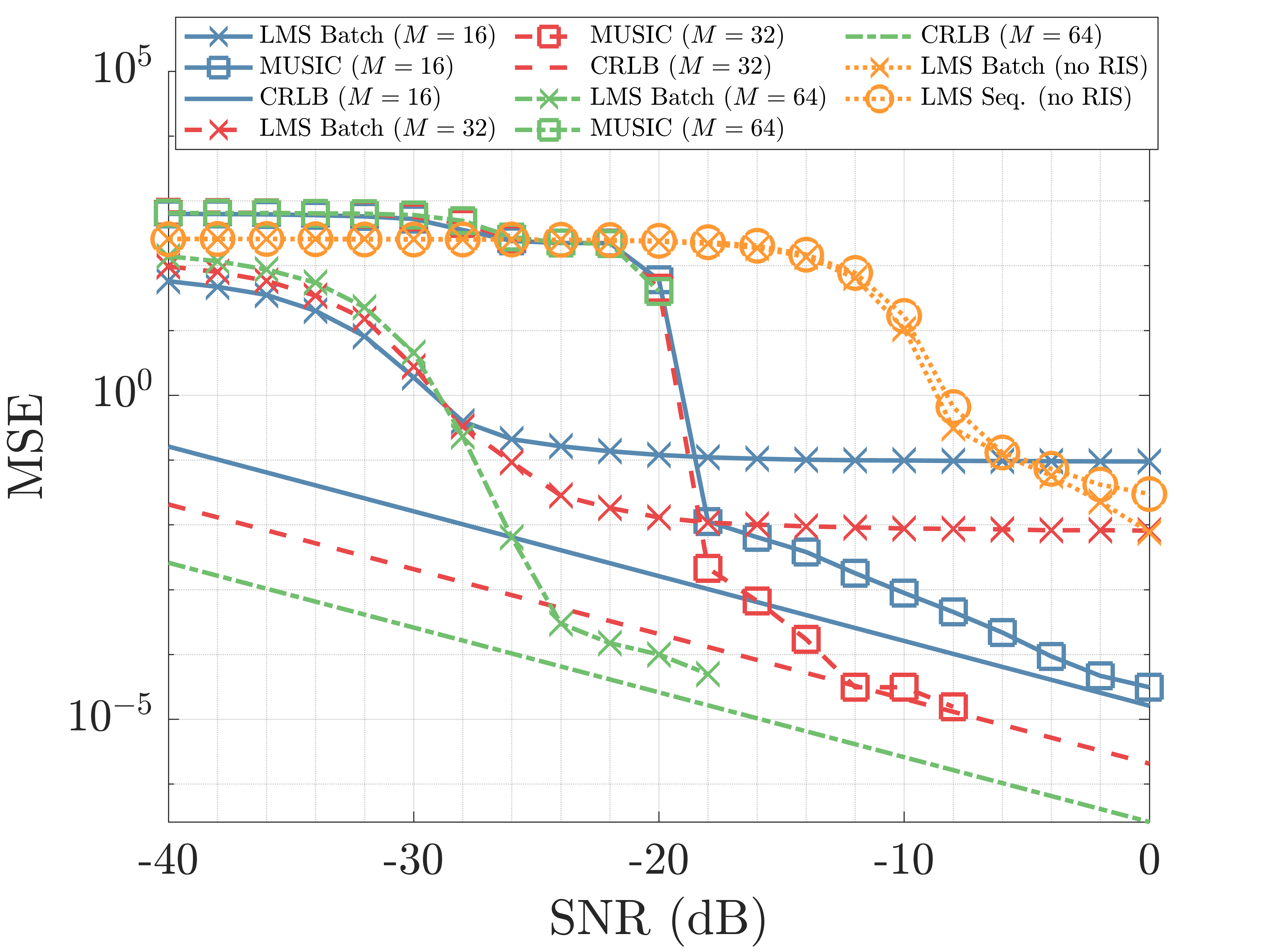}
	\caption{The \ac{MSE} vs. \ac{SNR} performances of batch \ac{NLMS}, sequential \ac{NLMS}, MUSIC-based estimation and \ac{CRLB}, with $K = 2$ targets, for \ac{ULA} configuration of $N_{\tt{PR}} = 8$, $L = 100$, and different values of $M$.}
	\label{fig:MSEvsSNR}
\end{figure}

\begin{figure}[t!]
	\centering
	\includegraphics[width=0.8\linewidth]{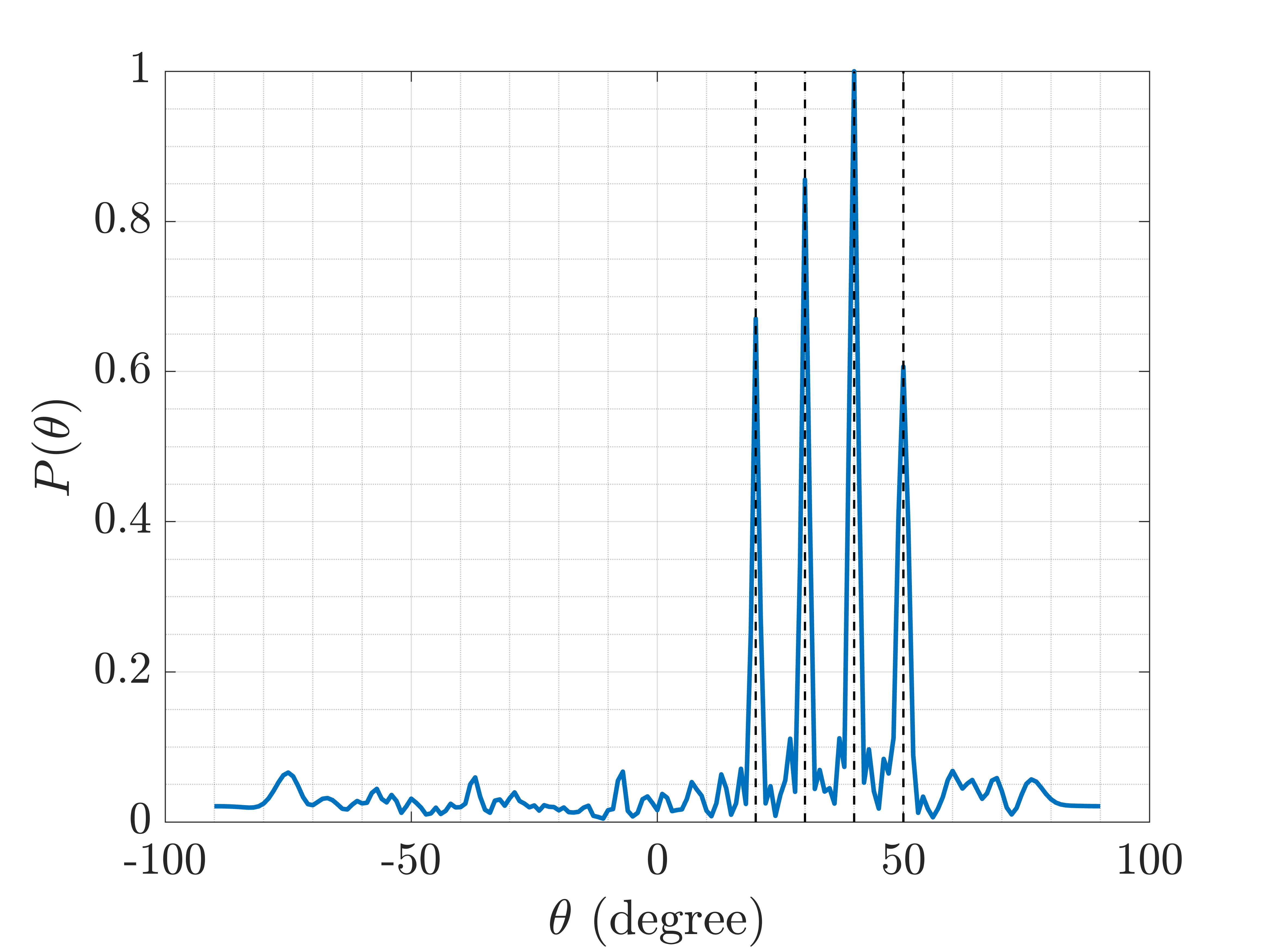}
	\caption{The spectrum of the proposed \ac{NLMS} method utilizing the \ac{RIS}-aided architecture. The scene consists of $K = 4$ targets located at $\theta_k^{\tt{RIS}} = 20 + 10(k-1), \forall k$. The \ac{RIS} is equipped with $M = 64$ reflective elements and the number of time samples is $L = 100$. The number of epochs is set to $N_{\tt{epoch}} = 100$. Dashed vertical lines represent the true \acp{AoA}.}
	\label{fig:spectrum-LMS-batch}
\end{figure}

\begin{figure}[t!]
	\centering
	\includegraphics[width=0.8\linewidth]{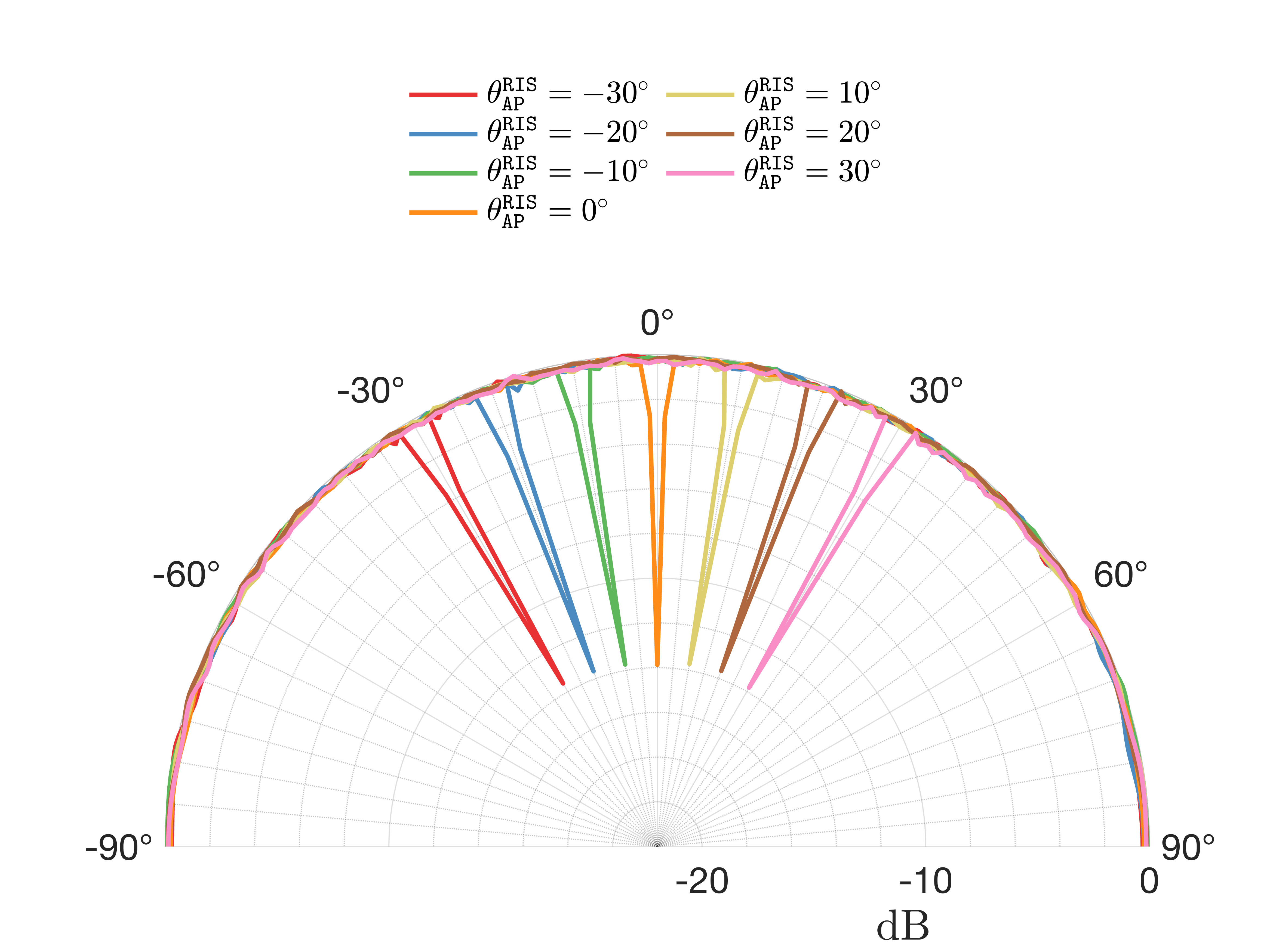}
	\caption{Normalized \ac{RIS} beampattern for different AP positions.}
	\label{fig:RISSelective}
\end{figure}

In this section, we showcase various simulation outcomes designed to demonstrate the efficacy of the proposed approaches. Before, we introduce the parameter configuration and elucidate the benchmarks employed for the simulation.
The number of time samples is set to $L=100$. 
Additionally, the \ac{PR} is equipped with $N_{\tt{PR}} = 8$ antenna elements. 
The peak detection threshold, utilized in both batch and sequential \ac{NLMS} method outlined in {{\textbf{Algorithm \ref{alg:cap}}}}, is set to $\varphi = 0.5$. 
The \acp{AoA} between the AP and the \ac{RIS}, as well as the \ac{RIS} and the \ac{PR}, are fixed at $\theta_{\tt{AP}}^{\tt{RIS}} = -10^\circ$ and $\theta_{\tt{RIS}}^{\tt{PR}} = -40^\circ$, respectively. 
A \ac{ULA} configuration with half-wavelength spacing is employed at the \ac{PR}. Unless specified otherwise, the number of epochs is maintained at $N_{\tt{epoch}} = 100$.
To facilitate performance comparisons, our performance study incorporates the following benchmarks:
\begin{itemize}
	\item \textbf{No-RIS Assistance:} In this scenario, the benchmark does not accommodate a \ac{RIS} for redirecting bounces from the targets. Without the \ac{RIS}, this benchmark lacks the ability to beamform towards a \ac{RIS}, requiring the \ac{PR} to estimate the target \ac{AoA}s embedded within $\pmb{h}_{k}^{\tt{PR}}$. Therefore, the \ac{PR} in this case observes
\begin{equation}
\label{eq:Rx-data-at-PR-on-RIS}
\begin{split}
	\pmb{Y}	
	= 
	\rho_{\tt{AP}}^{\tt{PR}} 
	\pmb{h}_{\tt{AP}}^{\tt{PR}}
	\pmb{s}_{\tt{AP}}^{\tt{PR}}
	+
	\sum\nolimits_{k=1}^K
	\rho_k
	\pmb{h}_{k}^{\tt{PR}}
	\pmb{s}_k^{\tt{PR}}
	+
	\pmb{E}_n.
\end{split}
\end{equation}
For this benchmark, {{\textbf{Algorithm \ref{alg:cap}}}} can be repurposed. In this case, the algorithm estimates $\theta_k^{\tt{PR}}$ instead of $\theta_k^{\tt{RIS}}$ due to the absence of an \ac{RIS}. Specifically, the input becomes $\pmb{Y}$, and line $6$ is replaced with $p_{\ell}(\theta)
\gets
\pmb{a}^H(\theta)
\pmb{y}_\ell$ since there is no reflective coefficient matrix $\pmb{V}$. Additionally, line $7$ is modified to $ \hat{\pmb{a}}_{\ell + 1}(\theta)
\gets
\hat{\pmb{a}}_{\ell}(\theta)
+
\frac{\mu}{\Vert \pmb{z}_{\ell} \Vert} ( p_{\ell}(\theta) - \hat{\pmb{a}}_{\ell}^H(\theta) \pmb{y}_\ell )^* \pmb{y}_\ell $.
\item \textbf{\ac{CRLB} Benchmark}: We also include the \ac{CRLB} benchmark, which is not derived in this paper due to lack of space. Indeed, this benchmark aims at quantifying how good an estimator is operating as compared to the best un-biased estimator of target \ac{AoA}s via the proposed \ac{RIS}-aided architecture for target localization.
\item \textbf{MUSIC-based Estimation with complete knowledge of number of targets}: For benchmarking, we also design a \ac{MUSIC} estimator tailored for estimating target \ac{AoA}s at the \ac{PR} using the proposed architecture. 
This estimator is aware of the true number of targets in the environment, which is further used to correctly separate signal and noise subspaces.
\end{itemize}
In Fig. \ref{fig:MSEvsSNR}, we aim at analyzing the \ac{MSE} as a function of \ac{SNR} at the \ac{PR}. The \ac{MSE} is computed as 
\begin{equation}
	\MSE = \frac{1}{PK} \sum\nolimits_{k=1}^K \sum\nolimits_{p = 1}^{P} (\theta_k - \hat{\theta}_k)^2,
\end{equation}
The parameter $\theta_k$ is the true \ac{AoA} of the $k^{th}$ target and $\hat{\theta}_k$ is the estimated \ac{AoA} of the $k^{th}$ target. We have simulated two targets, i.e. $K = 2$. Also,  $P$ is the number of Monte-carlo trials. 
We aim at showing the performance for increasing number of reflective elements and for different benchmarks.
First, we note that the batch \ac{NLMS} method approaches the \ac{CRLB} upon increasing $M$.
Focusing on $M = 16$ elements, we also observe that at a target \ac{MSE} of $0.2$, the batch \ac{NLMS} algorithm requires about $\SNR = -26\dB$, whereas the \ac{MUSIC}-based method requires an additional $5.75\dB$ compared to the \ac{NLMS} algorithm.
At this \ac{MSE} level, a gain superior to $20\dB$ can be attained.
Beyond $-18\dB$, we see that \ac{MUSIC} dominates the \ac{NLMS} methods, as \ac{MUSIC} exploits the knowledge of the exact number of targets through proper segregation of signal and noise subspaces.

Fig. \ref{fig:spectrum-LMS-batch} shows the spectrum obtained by the proposed \ac{NLMS} filtering methods. It is clear from the simulation that the peaks are sharp enough to successfully resolve the $K = 4$ targets. The low noise level is low and the   detected peaks can be relied upon to detect the existence of targets at the locations, because the peak values surpass the threshold $\varphi = 0.5$.

In Fig. \ref{fig:RISSelective}, we study the normalized beampattern at the \ac{RIS} for different placements of the AP with respect to the \ac{RIS}, i.e. by varying $\theta_{\tt{AP}}^{\tt{RIS}}$.
This gives an idea regarding the \ac{DR} reduction that can be attained.
In this simulation, we set $N_{PR} = 16$ antennas, with $M = 64$ elements and $L = 90$ samples. 
In particular, we solve for the phase shifts and set the phase shifts in $\pmb{\Theta}^{(n)}$ as given by problem $(\mathcal{P}_{\tt{RIS}})$, i.e. we configure $\pmb{\Theta}^{(n)}$ following equation \eqref{eq:beta-optimal-vals}. Next, we plot the beampattern at the \ac{RIS}.
Indeed, according to equation \eqref{eq:RIS-model-with-Theta}, the beampattern reads
\begin{equation*}
	B(\theta)
	=
	\sum\nolimits_n
	\pmb{a}^H_M(\theta)
	(\pmb{\Theta}^{(n)})^H 
	\pmb{b}_{M}^*(\phi_{\tt{RIS}}^{\tt{PR}})
	\pmb{b}_{M}^T(\phi_{\tt{RIS}}^{\tt{PR}})
	\pmb{\Theta}^{(n)} 
	\pmb{a}_M(\theta).
\end{equation*}
The pattern suggest that for different values of $\theta_{\tt{AP}}^{\tt{RIS}}$, the \ac{RIS} is successfully able to redirect the target bounces at around $0\dB$ (quasi-distortionless), while attenuating the channel between the \ac{AP} and itself by about $14\dB$. This means that the \ac{ADC} at the \ac{PR} can operate with a reduced \ac{DR} of $14\dB$.

%%
%%\vspace{-0.2cm}
\section{Conclusions and Future Work}
\label{sec:conclusions}
In short, this paper addresses the challenges faced by \ac{PR} in scenarios with weak \ac{LoS} signals, particularly those caused by communication signals from an \ac{AP} and further reflected off targets. 
The deployment of \ac{RIS} with strong \ac{LoS} coverage between itself and the \ac{AP}, as well as the \ac{PR}, is proposed to overcome the limitations of weak \ac{LoS} signals. 
The key contributions of this work include the formulation of a comprehensive system model, the introduction of a collaborative RIS-PR beamforming strategy to enhance LoS signal reception for localization, while keeping a reasonably low \ac{DR} at the \ac{PR} for \ac{ADC} power consumption purposes, and the presentation of an adaptive approach based on \ac{NLMS} for joint target number detection and \ac{AoA} estimation. Simulation results showcase the effectiveness of the proposed method, highlighting a significant $14\dB$ reduction in \ac{DR} achievable at the \ac{PR}. This research offers valuable insights into improving the performance of \ac{PR} systems in challenging communication environments.
Future work will focus on designing and solving optimization frameworks that can achieve a lower \ac{DR}, while maintaining a good localization and detection performance.

\bibliographystyle{IEEEtran}
\bibliography{refs}

\end{document}